\documentclass[preprint,nofootinbib]{revtex4-1}

\usepackage{latexsym,amsmath,amssymb,graphicx,epsfig}
\usepackage{multirow,float,url,subcaption,xcolor,ulem}

\begin{document}

\title{Construction of $bb\bar{u}\bar{d}$ tetraquark states on lattice \\
with NRQCD bottom and HISQ up/down quarks}

\author{Protick Mohanta}
\author{Subhasish Basak}
\affiliation{School of Physical Sciences, National Institute of
Science Education and Research, HBNI, Odisha 752050, India}

\date{\today}

\begin{abstract}
We construct $bb\bar{u}\bar{d}$ states on lattice using NRQCD
action for bottom and HISQ action for the light up/down quarks.
The NRQCD-HISQ tetraquark operators are constructed for
``bound" $[bb][\bar{u}\bar{d}]$ and ``molecular" $[b\bar{u}]
[b\bar{d}]$ states. Corresponding to these different operators,
two different appropriately tuned light quark masses are needed
to obtain the desired spectra. We explain this requirement of
different $m_{u/d}$ in the light of relativised quark model
involving Hartree-Fock calculation. The mass spectra of double
bottom tetraquark states are obtained on MILC $N_f=2+1$ Asqtad
lattices at three different lattice spacings. Variational analysis
has been carried out to obtain the relative contribution of ``bound"
and ``molecular" states to the energy eigenstates.
\end{abstract}

\maketitle

\section{Introduction} \label{sec_intro}
The multiquark hadronic states other than the mesons and baryons are
relatively new entrants particularly in the heavy quark sector. The
signature of some of such states containing four or more quarks and/or
antiquarks have been observed in experiments \cite{bondar1, bondar2, aaij,
lebed, esposito, olsen}. The QCD states composed of
four valence quarks are popularly referred to as tetraquark, which is
used to denote either bound or often both bound and two 2-quark mesonic
particles bound in a molecular like structure. In this paper we use
the term tetraquark in the latter sense. Such states are characterized
by $J^{PC}$ quantum numbers that cannot be arrived at from the quark
model. However, heavy hadronic tetraquark states $QQ\bar{l} \bar{l}$
and their stability in the infinite quark mass limit had been studied
in \cite{manohar, eichten} which raised the possibility of existence
of heavy four quark bound states below the $Q\bar{l} - Q\bar{l}$
threshold. Of late, the observations of $Z^- (4430, \,1^+)$ of minimal
quark content being $c\bar{c}d\bar{u}$ \cite{aaij} have been reported
along with the $1^+$ states like $Z_b(10610)$ and $Z_b^\prime(10650)$,
having minimal quark content of four quarks (containing a $b\bar{b}$
pair) that are a few MeV above the thresholds of $B^\star \bar{B}\,
(10604.6)$ and $B^\star \bar{B}^\star\, (10650.2)$ \cite{bondar1,
bondar2}. The proximity of $Z_b,\, Z_b^\prime$ to the $B^\star
\bar{B}^\star$ threshold values perhaps suggest molecular, instead
of bound, nature of the states.

Around the same time, lattice QCD has been employed to investigate
the bound and/or molecular nature of the heavy tetraquark states,
not only to understand the above experimentally observed states
but also to identify other possible bound tetraquark states in both
$0^+$ and $1^+$ channels. In the context of lattice
study of heavy tetraquarks, some early lattice calculations in charm sector
involve $T_{cc}$ and $T_{cs}$ tetraquark states \cite{ikeda}, $cc\bar{c}
\bar{c}$ \cite{wagner}, $X(3872)$ and $Y(4140)$ \cite{padmanath}
and more recently $D_{s0}^\star (2317)$ \cite{alexandrou}. The bottom
sector too received intense attention in the past few
years. However, on the lattice the $Z_b,\,Z_b^\prime$ of quark content
$[b\bar{b}u\bar{d}]$ has been replaced with relatively simpler $[bb
\bar{u}\bar{d}]$ or equivalently $[\bar{b}\bar{b}ud]$ system; that is
to say instead of $B^*\bar{B}$ or $B^*\bar{B}^*$, the study is basically
on $BB^*\,/\,B^*B^*$ systems. One exploratory lattice study of the
$[b\bar{b} u\bar{d}]$ system has been reported in \cite{sasa}. The
lattice investigations this far
involve four bottom $bb\bar{b} \bar{b}$ \cite{hughes} and two bottom
tetraquark states $\bar{b}\bar{b} l_1 l_2$, where $l_1, l_2 \in c, s,
u, d$, \cite{francis1,francis2, junnarkar,luka}. An important
observation of these lattice studies is that the possibility of the
existence of $bb\bar{l}_1\bar{l}_2$ tetraquark bound states increases
with decreasing light quark masses, while they become less bound with
decreasing heavy (anti)quark mass.

Besides the usual lattice simulations, the heavy tetraquark systems
have also been studied using QCD potential \cite{richard}
and Born-Oppenheimer approximation \cite{bicudo1,bicudo2,bicudo3,bicudo4}.
The main idea in these references is to investigate tetraquark states with
two heavy (anti)quarks, which was $\bar{b}\bar{b}$ in the study, and two
lighter quarks using quantum mechanical Hamiltonian containing screened
Coulomb potential. This approach has been used to explain our two different
choices of light $u/d$ quark masses for different classes of tetraquark
operators.

In this work our goal is to construct tetraquark states, having quark
content $bb\bar{l}_1\bar{l}_2$ in $1^+$ both below and above $B-B^*$
threshold, by a combination of lattice operators and tuning quark
masses based on quantum mechanical potential calculation. For the
$b$ quark, we employed nonrelativistic QCD formulation \cite{thacker,
lepage}, as is the usual practice, and HISQ action \cite{hisq} for
$l_1, l_2 = u/d$. Here we also explore through variational/GEVP
analysis how the trial states created by our operators contribute
to the energy eigenstates.

First, we briefly review the salient features and parameters of both
NRQCD and HISQ actions along with the steps involve in combining the
relativistic $u/d$ HISQ propagators with the NRQCD $b$ quark
propagators in the section \ref{sec_action}. We have considered two
different kind of operators -- the local heavy diquark and light
antidiquark (often referred as ``good diquark" configuration) and
molecular meson-meson, we described these constructions in the
section \ref{sec_operator}. We collect our spectrum results in
section \ref{sec_numerical} that contains subsections on quark
mass tuning (\ref{subsec_tune}), Hartree-Fock calculation of two
light quarks in the presence of a heavy quark (\ref{subsec_hartree}),
tetraquark spectra (\ref{subsec_spectra}) and GEVP analysis
(\ref{subsec_gevp}). Finally we summarized our results in section
\ref{sec_summary}.

\section{Quark actions} \label{sec_action}
Lattice QCD simulations with quarks require quark mass to be $am_l
\ll 1$, where $a$ is the lattice spacing. In the units of the lattice
spacings presently available, the $b$ quark mass is not small {\it
i.e.} $am_b \nless 1$. As is generally believed, the typical velocity
of a $b$ quark inside a hadron is
nonrelativistic $v^2 \sim 0.1$ and is much smaller than the bottom
mass. This makes NRQCD our action of choice for $b$ quarks on lattice.
We have used $\mathcal{O}(v^6)$ NRQCD action \cite{lepage}, where the
Hamiltonian is $H = H_0 + \delta H$, where $H_0$ is the leading
$\mathcal{O}(v^2)$ term, the $\mathcal{O}(v^4)$ and $\mathcal{O}(v^6)$
terms are in $\delta H$ with coefficients $c_1$ through $c_7$,
\begin{eqnarray}
H_0 &=& -{\tilde{\Delta}^2 \over 2 m_b}-{a\over 4n}{(\Delta^2)^2\over
4 m_b^2} \label{nrqcd_h0} \\
\delta H &=& -c_1\, {(\Delta^2)^2\over 8 m_b^3} +
c_2\, {ig\over 8 m_b^2} \, \left(\vec{\Delta}^{\pm} \cdot \vec{E} -
\vec{E} \cdot \vec{\Delta}^{\pm} \right) - c_3\, {g\over 8 m_b^2}\;
\vec{\sigma} \cdot \left(\tilde{\vec{\Delta}}^{\pm}\times\tilde{\vec{E}}
- \tilde{\vec{E}}\times \tilde{\vec{\Delta}}^{\pm} \right) \nonumber \\
&& -c_4\, {g\over 2m_b}\; \vec{\sigma} \cdot \tilde{\vec{B}} -c_5\,
{g\over 8m_b^3}\, \left\{\Delta^2, \vec{\sigma} \cdot \vec{B}\right\}
-c_6\, {3g\over 64m_b^4}\, \left\{\Delta^2,\vec{\sigma} \cdot
\left(\vec{\Delta}^{\pm}\times\vec{E}-\vec{E}\times \vec{\Delta}^{\pm}
\right) \right\} \nonumber \\
&& -c_7\, {ig^2\over 8m_b^3}\; \vec{\sigma} \cdot \vec{E} \times\vec{E}
\label{nrqcd_dh}
\end{eqnarray}
where $\Delta^\pm$ and $\Delta^2$ are discretized symmetric covariant
derivative and lattice Laplacian respectively. Both the derivatives
are $\mathcal{O}(a^4)$ improved as are the chromoelectric $\vec{E}$
and chromomagnetic $\vec{B}$ fields. The $b$ quark propagator is
generated by time evolution of the Hamiltonian $H$,
\begin{eqnarray}
G(\vec{x},t+1;0,0) &=& \left(1-{aH_0\over 2n}\right)^n \left(1-{a\delta
H\over 2}\right) U_4(\vec{x},t)^{\dagger} \left( 1 - {a\delta H
\over 2}\right) \left(1-{aH_0\over 2n}\right)^n G(\vec{x},t;0,0)
\nonumber \\
&& \label{green} \\
&& \text{with} \hspace{0.15in}
G(\vec{x},t;0,0)= \left\{ \begin{array}{cl} 0 & \hspace{0.15in} \text{for}
\;\; t<0 \\ \delta_{\vec{x},0} & \hspace{0.15in} \text{for} \;\;
t=0 \end{array}\right. \nonumber
\end{eqnarray}
The tree level value of all the coefficients $c_1$, $c_2$, $c_3$, $c_4$,
$c_5$, $c_6$ and $c_7$ is 1. Here $n$ is the factor introduced to ensure
numerical stability at small $am_b$, where $n > 3/2m_b$ \cite{thacker}.

In NRQCD, the rest mass term does not appear either in the equation
(\ref{nrqcd_h0}) or in (\ref{nrqcd_dh}), and therefore, hadron masses
cannot be determined from their energies at zero momentum directly
from the exponential fall-off of their correlation functions. Instead,
we calculate the kinetic mass $M_k$ of heavy-heavy mesons from its
energy-momentum relation, which to $\mathcal{O}(p^2)$ is
\cite{fermilab},
\begin{equation}
E(p) = E(0) + \sqrt{p^2 + M_k^2} - M_k \;\;\;\Rightarrow \;\;\;
E(p)^2 = E(0)^2 + {E(0)\over M_k}\,p^2. \label{kin-mass}
\end{equation}
where $M_k$ is the kinetic mass of the meson which is calculated
considering $E(p)$ at different values of lattice momenta $\vec{p}
= 2\vec{n}\pi/L$. The $b$ quark mass is tuned from the spin average
of  kinetic masses of $\Upsilon$ and $\eta_b$, and matching them with
the experimental spin average value,
\begin{equation}
M_{b\bar{b}} = {3M_{\Upsilon} + M_{\eta_b}\over 4} \label{mbspinavg}
\end{equation}
The experimental value to which $M_{b\bar{b}}$ is tuned to, however,
is not 9443 MeV that is obtained from spin averaging $\Upsilon$ (9460
MeV) and $\eta_b$ (9391 MeV) experimental masses, but to an
appropriately adjusted value of 9450 MeV \cite{eric}, which we denote
as $M_{\text{phys}}^{\text{mod}}$ in the equation (\ref{mass_formula})
below. The hadron mass is then obtained from
\begin{equation}
M_{\text{latt}} = E_{\text{latt}} + {n_b\over2}\, \left(
M_{\text{phys}}^{\text{mod}} - E_{\text{latt}}^{\eta_b} \right)
\label{mass_formula}
\end{equation}
where $E_{\text{latt}}$ is the lattice zero momentum energy in MeV,
$n_b$ is the number of $b$-quarks in the bottom hadron.

The $u/d$ light quarks comfortably satisfy the criteria $am_l \ll 1$
and, therefore, we can use a relativistic lattice action. We use
HISQ action for the $u/d$ quarks, which is given in \cite{hisq},
\begin{equation}
\mathcal{S} = \sum_x \bar{q}(x) \left(\gamma^\mu D_\mu^{\text{HISQ}} +
m \right)\,q(x) \hspace{0.1in} \text{where,} \hspace{0.1in}
D_\mu^{\text{HISQ}} = \Delta_\mu(W) - {a^2\over 6}(1+\epsilon)\,
\Delta_\mu^3(x). \label{hisq_act}
\end{equation}
Because HISQ action reduces $\mathcal{O}(\alpha_s a^2)$ discretization
error found in Asqtad action, it is well suited for $u/d$ (and
$s$) quarks. The parameter $\epsilon$ in the coefficient of Naik
term can be appropriately tuned to use the action for $c$ quarks,
which we do not have here. For $u/d$ (and $s$) quarks, the
$\epsilon = 0$.

HISQ action is diagonal in spin space, and therefore, the
corresponding quark propagators do not have any spin structure.
The full $4 \times 4$ spin structure is regained by multiplying
the propagators by Kawamoto-Smit multiplicative phase factor
\cite{kawamoto},
\begin{equation}
\Omega(x) = \prod_{\mu=1}^4(\gamma_{\mu})^{x_{\mu}} = \gamma_{1}^{x_1}
\gamma_{2}^{x_2} \gamma_{3}^{x_3} \gamma_{4}^{x_4}.
\label{staggop}
\end{equation}

\section{tetraquark operators} \label{sec_operator}
In the present paper, we have considered two kinds of tetraquark
operators -- the local heavy diquark and light antidiquark and
molecular meson-meson. The $b$ quark, being nonrelativistic, is
expressed in terms of two component field $\psi_h$. We convert
it into a four component spinor $Q$ having vanishing two lower
components,
\begin{equation}
Q \equiv \begin{pmatrix} \psi_h \\ 0 \end{pmatrix} \label{Q_fld}
\end{equation}
which help us to combine $b$ field and relativistic four component
light quark fields in the usual way. The heavy-light meson operator,
that we will make use of in the operator construction, is
written as
\begin{equation}
\mathcal{O}_{hl}(x) = \bar{Q}(x)\, \Gamma\, l(x) \label{hlintpol}
\end{equation}
where $l(x)$ stands for the light quark fields, $\bar{Q} = Q^{\dagger}
\gamma_4$ and depending on pseudoscalar and vector mesons $\Gamma =
\gamma_5$ and $\gamma_i$ respectively.

Because of the vanishing lower components, the states with $Q$ can
only be projected to the positive parity states. The local double
bottom tetraquark operators that we can construct for $bb\bar{l}_1
\bar{l}_2$ system are,
\begin{align}
\mathcal{O}_{M_1} &\equiv \mathcal{O}_{B^*B} \;\,= \hspace{0.25in}
\left[ \bar{l}_1(x) \gamma_i Q(x) \right] \left[\bar{l}_2(x) \gamma_5
Q(x) \right], \label{bsb} \\
\mathcal{O}_{M_2} &\equiv \mathcal{O}_{B^*B^*} = \epsilon_{ijk}
\left[ \bar{l}_1(x) \gamma_j Q(x) \right] \left[\bar{l}_2(x) \gamma_k
Q(x)\right], \label{bsbs} \\
\mathcal{O}_D &\equiv \mathcal{O}_{\mathcal{Q}^*\tilde{\pi}} \;\;=
\hspace{0.27in} \left[ Q^{a\,T}(x)C\gamma_i Q^b(x) \right] \left[
\bar{l}^a_1(x) C\gamma_5 \bar{l}_2^{b\,T}(x) \right] \label{Qpi}
\end{align}
where $l_1\neq l_2$ and $l_1, l_2 \in u,d$. The
$a,\,b$ are the color indices. The naming convention above
is borrowed from reference \cite{francis2} but the exact construction
of the operators is different. In literature the operators in (\ref{bsb})
and (\ref{bsbs}) are often referred to as ``molecular".
In this work we have not included the non-local
operators in our GEVP analysis like those in \cite{luka}, although
they will admittedly effect the first few excited states.

The diquark-antidiquark $1^+$ four quark state $bb\bar{l}_1 \bar{l}_2$
with $l_1\neq l_2$ in (\ref{Qpi}) can actually be defined in two ways
\cite{ttqop},
\begin{eqnarray}
\mathcal{O}_{\mathcal{Q}^*\tilde{\pi}} &=& \left[Q^{a\,T} C\gamma_i \,
Q^b \right] \, \left[\bar{l}_1^a \,C\gamma_5\, \bar{l}_2^{b\,T} -\,
\bar{l}_1^b\,C\gamma_5 \,\bar{l}_2^{a \,T} \right] \nonumber \\
\mathcal{O}_{\mathcal{Q}\tilde{\pi}^*} &=& \left[Q^{a\,T}
C\gamma_5 \, Q^b \right] \, \left[\bar{l}_1^a \,C\gamma_i\,
\bar{l}_2^{b\,T} +\, \bar{l}_1^b\,C\gamma_i \,\bar{l}_2^{a
\,T} \right] \label{2D_ttq}
\end{eqnarray}
The subscripts $\mathcal{Q}^*$ and
$\tilde{\pi}$ in the operator $\mathcal{O}_{\mathcal{Q}^*\tilde{\pi}}$
are in $\bar{3}_c$ and $3_c$ respectively, while $\mathcal{Q}$ and
$\tilde{\pi}^*$ in the operator $\mathcal{O}_{\mathcal{Q}\tilde{\pi}^*}$
are in $6_c$ and $\bar{6}_c$. But both $\mathcal{O}_{\mathcal{Q}^*\tilde{
\pi}}$ and $\mathcal{O}_{\mathcal{Q}\tilde{\pi}^*}$ correspond to the
$1^+$ state. Of these the $\mathcal{O}_{\mathcal{Q}^*\tilde{\pi}}$ is
our desired ``bound" tetraquark operator because one-gluon-exchange
interaction is attractive for a heavy quark pair in $\bar{3}_c$
diquark configuration \cite{eichten} and spin dependent attraction
exists for light quark pairs in ``good diquark'' configuration
characterized by color $\bar{3}_c$, spin $J=0$ and isospin $I=0
\text{ or } 1/2$ \cite{jaffe}. The two terms in $\mathcal{O}_{
\mathcal{Q}^*\tilde{\pi}}$ contribute identically in the final
correlator, hence we consider only the first term in the calculation.
The generic form of the temporal correlation among the operators
at zero momentum is,
\begin{equation}
C_{XY}(t) = \sum_{\mathbf{x}} \left\langle[ \mathcal{O}_{X}(
\mathbf{x},t)]\, [\mathcal{O}_{Y}(\mathbf{0},0)]^\dagger \right
\rangle, \label{op_corr}
\end{equation}
where $X,Y$ can be any of $D, M_1, M_2$ in equations (\ref{bsb}),
(\ref{bsbs}) and (\ref{Qpi}). For example, the explicit forms of
the zero momentum correlators, including cross-correlator, when
$X$ and $Y$ are $M_1=B^* B$ and $D=\mathcal{Q}^* \tilde{\pi}$,
are
\begin{eqnarray}
C_{M_1\,M_1}(t) &=&
\sum_{\vec{x}}\text{Tr}\left[\gamma_5 \,M_1^{\dagger}(x,0)\,\gamma_5 \,\, 
\gamma_i\, G(x,0)\, \gamma_i \right] \times 
\text{Tr}\left[M_2^{\dagger}(x,0)\,\, G(x,0) \right] \nonumber \\
&& - \sum_{\vec{x}}\text{Tr}\left[G(x,0) \,M_2^{\dagger}
(x,0)\,G(x,0) \,\, \gamma_i\gamma_5\, M_1^{\dagger}(x,0)\,\gamma_5\gamma_i
\right] \label{BsB_cor} \\
C_{D\,D}(t) &=&
\sum_{\vec{x}} \text{Tr} \left[\left(G^{ad}(x,0)\right)^T \gamma_i
\gamma_4 \gamma_2\, G^{bc}(x,0) \gamma_4\gamma_2\gamma_i \right] \times
\nonumber \\
&& \hspace{0.3in} \text{Tr} \left[ \gamma_4\gamma_2 \,{M_1^\dagger}^{da}
(x,0)\, \gamma_4\gamma_2\,\left(\gamma_5{M_2^\dagger}^{cb}(x,0)\gamma_5
\right)^T \right] \label{QB_cor} \\
C_{D\,M_1}(t) &=& 
\sum_{\vec{x}} \text{Tr} \left[ G^{ad}(x,0)\,\gamma_i\gamma_5\,
{M_1^\dagger}^{da} (x,0)\, \gamma_5\gamma_2\gamma_4\gamma_5
\right]_{\mu\nu} \times \nonumber \\
&& \hspace{0.3in} \text{Tr} \left[ \gamma_2\gamma_i\gamma_4\,G^{bc}(x,0)
\, \gamma_5\,\, \gamma_5\, {M_2^\dagger}^{cb}(x,0)\gamma_5
\right]_{\mu\nu} \label{QBsB_cor}
\end{eqnarray}
Above in the equations (\ref{QB_cor}) and (\ref{QBsB_cor}), traces and
transposes are taken over the spinor indices,  while in the equation
(\ref{BsB_cor}) the traces are taken over both the spinor and color indices.
Here $G(x,0)$ denotes the heavy quark propagators while the $M(x,0)$ are
the light quark propagators. The term Tr$\,\left[ G
M_2^\dagger GM_1^\dagger \right]$ appears only in $C_{M_1 M_1}$,
$C_{M_2M_2}$, $C_{M_1M_2}$ and $C_{M_2M_1}$ correlators. The diquark and the
anti-diquark part of $\mathcal{O}_D$ in (\ref{Qpi}) do not have free
spinor index and, therefore, we do not have similar Tr$\,\left[G
M_2^\dagger GM_1^\dagger \right]$ term in $C_{D\,D}$.  The remaining
correlators $C_{D\,M_1}$, $C_{D\,M_2}$, $C_{M_1\,D}$ and $C_{M_2\,D}$
can not be expressed in compact Tr$\,\left[G\,M_1^\dagger\right]\times$
Tr$\,\left[G,M_2^\dagger \right]$ form. In terms of coding, we have
to keep in mind that NRQCD and MILC library suit use different
representation of gamma matrices. Therefore the heavy quark propagator
$G(x,0)$ has to be rotated to the MILC basis before implementing the
equations (\ref{BsB_cor}), (\ref{QB_cor}) and (\ref{QBsB_cor}). The
unitary matrix needed to do this transformation is given by \cite{protick},
\begin{equation}
S \, \gamma^{\text{MILC}}_{\mu}\, S^{\dagger} = \gamma^{\text{NR}}_{\mu}
\hspace{0.1in} \text{where,} \hspace{0.1in} S = {1\over \sqrt{2}}
\begin{pmatrix}
  \sigma_y && \sigma_y\\
  -\sigma_y && \sigma_y
   \end{pmatrix}. \label{smatrix} 
\end{equation}

\section{Numerical Studies} \label{sec_numerical}
We calculated the double bottom tetraquark spectra using the publicly
available $N_f=2+1$ Asqtad gauge configurations generated by MILC
collaboration. Details about these lattices can be found in
\cite{bazavov}. It uses Symanzik-improved L\"{u}scher-Weisz action
for the gluons and Asqtad action \cite{asqtad1, asqtad2} for the
sea quarks. The lattices we choose have a fixed ratio of $am_l/am_s
=1/5$ with lattice spacings 0.15 fm, 0.12 fm and 0.09 fm and they
correspond to the same physical volume. We have not determined the
lattice spacings independently but use those given in \cite{bazavov}.
In the Table \ref{tab:milclatt} we listed the ensembles used in this
work.
\begin{table}[h]
\caption{MILC $N_f=2+1$ Asqtad configurations used in this work. The
gauge coupling is $\beta$ and the lattice spacing is $a$. The $u/d$
and $s$ sea quark masses are $m_l$ and $m_s$ respectively and the
lattice size is $L^3\times T$. The $N_{\text{cfg}}$ is number of
configurations used in this work.}
\begin{ruledtabular}
\begin{tabular}{cccccc}
$\beta={10/ g^2}$ & $a$(fm) & $am_l$ & $am_s$ & $L^3\times T$ &
$N_{\text{cfg}}$    \\       
\hline
6.572 & 0.15 & 0.0097 & 0.0484 & $16^3\times 48$ & 600 \\
6.76 & 0.12 & 0.0100 & 0.0500 & $20^3\times 64$ & 600 \\
7.09 & 0.09 & 0.0062 & 0.0310 & $28^3\times 96$ & 300 \\
\end{tabular}
\end{ruledtabular}
\label{tab:milclatt}
\end{table}

\subsection{Quark mass tuning} \label{subsec_tune}
For $bb\bar{u}\bar{d}$ mass calculation, we need nonperturbative tuning
of both $m_b$ and $m_{u/d}$. With the help of equation (\ref{mbspinavg}),
the tuning of $m_b$ has been carried out by calculating the spin average
of $\Upsilon$ and $\eta_b$ kinetic masses and comparing the same with the
spin average and suitably adjusted experimental $\Upsilon$ and $\eta_b$
masses as discussed in section \ref{sec_action}. The tuned bare $am_b$
quark masses for lattices used in this work are given in Table
\ref{tab:qmass}.
\begin{table}[h]
\caption{Tuned $b$ and $u/d$ quark bare masses for lattices used in
this work. For $u/d$-quark mass, we mention the particle states used
to tune.}
\begin{ruledtabular}
\begin{tabular}{llccc}
\multirow{2}{*}{Quark} & Tuning & $16^3\times 48$ & $20^3\times 64$
 & $28^3\times 96$ \\
 & hadron & (0.15 fm) & (0.12 fm) & (0.09 fm) \\ \hline
$am_b$ & $\Upsilon - \eta_b$ & 2.76 & 2.08 & 1.20 \\
$am_{u/d}$ & $\Lambda_b$ (5620) & 0.105 & 0.083 & 0.064 \\
$am_{u/d}$ & $B$ (5280) & 0.155 & 0.118 & 0.087 \\
\end{tabular}
\end{ruledtabular}
\label{tab:qmass}
\end{table}

But the tuning of $am_{u/d}$ is rather tricky. It was
found in \cite{protick} that the $B$-meson tuned $am_{u/d}$ reproduces
the mass of $\Sigma_b$ baryon but not that of $\Lambda_b$. Therefore,
we tuned $am_{u/d}$ to two different values depending on the construction
of the pairs $[ud]$ and $[bu]$ or $[bd]$. The motivation to do so followed
from the observation that substantial mass difference exist among singly
heavy baryons having same quark content and same $J^P$. For instance,
the mass differences between the $J^P={1\over 2}^+$ pairs $(\Lambda_b,\,
\Sigma_b \,[bdu])$, $(\Lambda_c, \,\Sigma_c \,[cdu])$, $(\Xi_b, \,
\Xi_b^\prime \,[bsu])$ and $(\Xi_c, \,\Xi_c^\prime \,[csu])$ are in the
range $110 - 190$ MeV. The $\Lambda_b$, $\Lambda_c$, $\Xi_b$ and
$\Xi_c$ baryons are characterized by the spin of the $[l_1 l_2]$ (where
$l_{1,2} \in u,\,d,\,s$) light-light diquark $s_l = 0$ while $\Sigma_b$,
$\Sigma_c$, $\Xi_b^\prime$ and $\Xi_c^\prime$ by $s_l=1$. This differences
in their wave functions alone cannot generate such mass differences
\cite{bowler_54} but can at most account for a difference of about 30 MeV.
The heavy hadron chiral perturbation theory calculations \cite{tiburzi,
zachary} for $\Lambda_Q$ and $\Sigma_Q$, where $Q \in b,c$, demonstrated
that the mass differences get large correction of the order $\approx 150$
MeV. A correction of similar magnitude is anticipated in our NRQCD-HISQ
heavy baryon / tetraquark systems, but the relevant $\chi$PT for which
is yet to be available. To include such a correction in our calculations,
we propose this unique method of tuning the $[\bar{u} \bar{d}]$ diquark
system to $\Lambda_b$-baryon and $[b \bar{u}]$ to $B$-meson.

In the present paper, we try to understand this tuning scheme in more
details with the help of relativised quark model \cite{mes_q_mod,bar_q_mod}
and Hartree-Fock calculation. The basic idea is that $am_{u/d}$ has to
be tuned to two different values corresponding to two different
constructions of the pairs $[\bar{u} \bar{d}]$ and $[b\bar{u}]$. In the
operator $\mathcal{O}_D \equiv [bb][\bar{u}\bar{d}]$, the anti-diquark
part formed with two light $u/d$ quarks is the same that appear in the
baryonic operator $\Lambda_b \equiv (u^TC\gamma_5 d)\,b$, and hence, we
use experimental $\Lambda_b$ mass 5620 MeV to tune the bare $am_{u/d}$.
For the operators $\mathcal{O}_{M_1/M_2}$, the diquark part is formed
between heavy quark and light antiquark $[b\bar{u}]$ which is the same
as in the $B$-meson $(\bar{b}\gamma_{(5,k)}u)$ or $\Sigma_b \equiv
(Q^T C\gamma_5\, u)\,u$. In such case we tend to use $B$-meson mass
5279 MeV to tune the $am_{u/d}$.

For $\mathcal{O}_D$ we have exclusively used $am_{u/d}$
tuned from $\Lambda_b$ where as for determining the lattice tetraquark
thresholds $B-B^*$ it is $B$ tuned. But $\mathcal{O}_D,\,
\mathcal{O}_{M_1},\,\mathcal{O}_{M_2}$ all have the
same quantum numbers and, therefore, expected to mix and contribute
to the finite volume lattice ``bound" states. When looking for such
bound states, we will be using $\Lambda_b$ tuned $am_{u/d}$ for all
the operators. However, while searching for purely molecular states,
in light of Hartree-Fock calculation in section \ref{subsec_hartree}
below we entirely omit $\mathcal{O}_D$ and worked only with the
$\mathcal{O}_{M_1/M_2}$ operators using just the $B$ tuning.

This differently tuned $am_{u/d}$ gave consistent result
in \cite{protick} and expect to repeat the same in this work. The
results of $u/d$ quark mass tuning that are made use of in this work
is given in the Table \ref{tab:qmass}.

Before we calculate the spectra of the $|D\rangle,\,|M_1\rangle,\,
|M_2\rangle$ tetraquark states defined in subsection \ref{subsec_spectra},
in the following subsection \ref{subsec_hartree} we try to understand
the diquark dependent different tuning of the light $u/d$ quark masses.
For this we consider Schr\"{o}dinger Hamiltonian for ($a$) hydrogen-like
system, namely $B$ meson with an $\bar{u}$ antiquark in the potential
of a static $b$ quark and ($b$) helium-like system, which is $\Lambda_b$
baryon with $u,\,d$ quarks in the same $b$ quark field.

\subsection{Hartree-Fock calculation of tetraquark states} \label{subsec_hartree}

\begin{center}
\begin{figure}[h]
\includegraphics[scale=0.9]{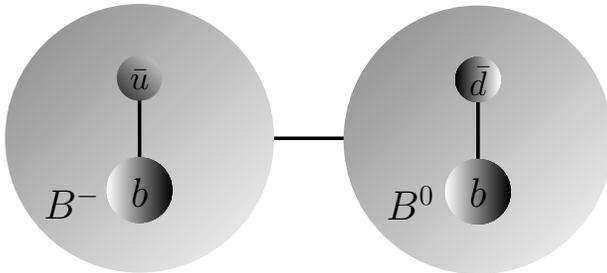}
\caption{Molecular tetraquark state viewed as bound state of two $B$
mesons, which is similar to two hydrogen atoms forming a hydrogen
molecule.}
\label{fig:molecule}
\end{figure}
\end{center}

\begin{center}
\begin{figure}[h]
\begin{subfigure}[b]{0.45\textwidth}
\includegraphics[scale=0.8]{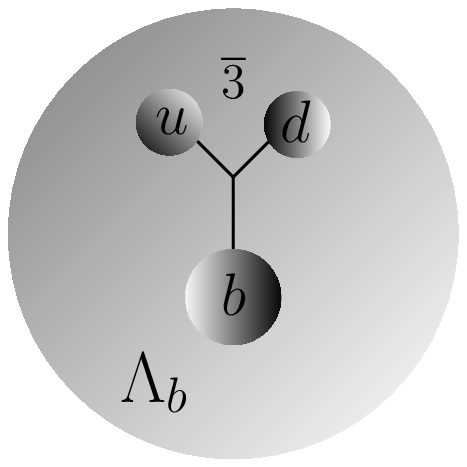}
\caption{$u/d$ quarks in $\Lambda_b$ baryons form a $\overline{3}_c$
diquark in presence of a $b$ quark.}
\label{fig:lambda}
\end{subfigure}
\begin{subfigure}[b]{0.45\textwidth}
\includegraphics[scale=0.8]{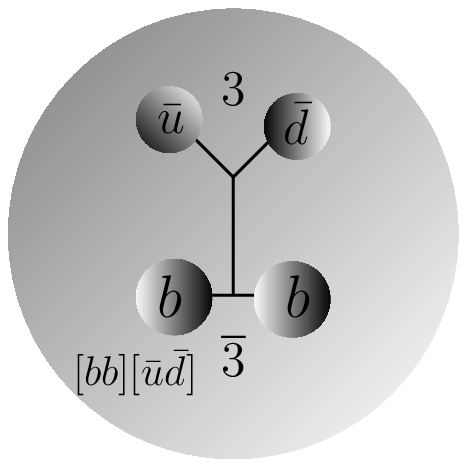}
\caption{Like $\Lambda_b$, two $b$ quarks form a (nearly) static
nucleus surrounded by $\bar{u},\,\bar{d}$ cloud.}
\label{fig:tetraquark}
\end{subfigure}
\caption{Schematic diagram of helium-like $\Lambda_b$ and $[bb][\bar{u}
\bar{d}]$ tetraquark state used for Hartree-Fock treatment.}
\label{fig:lambda_4q}
\end{figure}
\end{center}

In order to gain a qualitative understanding of two different tunings of
$m_{u/d}$, we consider the light antiquark and light-light diquark in the
potential of heavy, nearly static color source, the $b$ quark(s). This
picture is akin to hydrogen and helium-like quantum mechanical systems.
For the molecular tetraquark states, the basic assumption is that the
light antiquark wave functions do not have significant overlap with
each other and they are effectively in the potential of their respective
heavy $b$ quarks \cite{bicudo2} i.e. a two $B$-meson like system. But
for the  diquark-antidiquark tetraquark state $[bb][\bar{u}\bar{d}]$
where antidiquark component is similar to the $\Lambda_b$ light-light
diquark, we tune the $u/d$ quark mass using the $\Lambda_b$ baryon. The
situation is depicted schematically in Figs. \ref{fig:molecule}
and \ref{fig:lambda_4q}. The relevant interpolating operators for
$\Lambda_b$ and $B$ meson are fairly standard but for HISQ light
quarks a whole array of bottom baryon operators, including
$\Lambda_b$ can be found in \cite{protick}.

The relativised quark model \cite{mes_q_mod, bar_q_mod} helps us to
numerically calculate the masses of $B$ meson and $\Lambda_b$ baryon
using the light (anti)quark mass as parameter. The molecular
tetraquark state can be visualized as two $B$ meson molecule as shown
in the Fig. \ref{fig:molecule}. Then for each $B$ meson, the light
$u/d$ antiquark is taken to be in the field of ``static" $b$ quark
and we solve the problem by considering the radial part of the
Schr\"{o}dinger equation numerically using suitably modified
Herman-Skillman code \cite{herman}.
\begin{equation}
-{1\over 2m_{u/d}}{d^2U(r)\over dr^2} + V(r)U(r) = E U(r)
\label{schro_eq}
\end{equation}
Here $U(r) = r\psi(r)$ and the potential $V(r)$ is given by
\begin{equation}
 V(r)= -{4\alpha\over 3r}+\beta r \label{meson_pot}
\end{equation}
The $B$ meson mass $M_B$ is, therefore, determined from the energy
eigenvalue $E$,
\begin{equation}
M_B = m_b + m_{u/d} + E \label{mass_mb}
\end{equation}
where $m_b = 4.18$ GeV ($\overline{\text{MS}}$) is the mass of the
bottom quark, the $\alpha = \pi/16$ \cite{bicudo3} and $\beta = 0.18$
GeV\textsuperscript{2} \cite{mes_q_mod}. For $M_B = 5.279$ GeV, the
light quark mass obtained is $m_{u/d} \approx 0.227$ GeV.

For $\Lambda_b$ baryon, we used Hartree-Fock method
\cite{hartree1,hartree2} to solve the helium-like Hamiltonian,
\begin{eqnarray}
H &=& - {1\over 2m_{u/d}}\nabla_1^2 - {2\alpha\over 3r_1} + {\beta r_1
\over 2} - {1\over 2m_{u/d}}\nabla_2^2 - {2\alpha\over 3r_2} + {\beta
r_2 \over 2} \nonumber\\
&& - {2\alpha^\prime\over 3r_{12}} + {\beta^\prime r_{12} \over 2}
\label{hartree}
\end{eqnarray}
where $r_{12}$ is the relative distance between two light quarks 
``orbiting" the heavy quark and their interaction potential is
the last two terms in the equation (\ref{hartree}) with coefficient
$\alpha^\prime$ and $\beta^\prime$. For the Hartree-Fock calculation
of the energy $E$, we take $\beta^\prime = \beta$ and $\alpha^\prime
= 0.6$  \cite{mes_q_mod}.

To solve the Hamiltonian (\ref{hartree}), we consider the trial wave
function, which is space-symmetric and spin-antisymmetric, in terms
of Slater determinant
\begin{equation}
\Psi^{\textmd{HF}}={1\over \sqrt{2}} {\begin{vmatrix} \chi_1(x_1) &
\chi_1(x_2) \\ \chi_2(x_1) & \chi_2(x_2)\end{vmatrix}}, \label{slater}
\end{equation}
where $x_i \equiv (\vec{r},s)$ collectively denotes the space and spin
indices,  $\chi_i(\vec{r},s)=\phi_{i\,s}(\vec{r})\,\mathcal{S}(s)$ with
$\phi(\vec{r})$ being the 1S state. Therefore, the expectation value of
the the Hamiltonian can be written as
\begin{eqnarray}
\langle \Psi^{\textmd{HF}} | H |\Psi^{\textmd{HF}}\rangle &=&
\langle T \rangle + \int \rho(\vec{r})\,V_{\textmd{ext}}(\vec{r})
\,d\vec{r} - {Z^\prime \over 2} \iint {\rho(\vec{r})\rho(\vec{r}_1)\over
|\vec{r}-\vec{r}_1|}\, d\vec{r} \, d\vec{r}_1\nonumber\\
 && +\; {B^\prime \over 2} \iint {\rho(\vec{r})\,\rho(\vec{r}_1)\,
|\vec{r}-\vec{r}_1|} \,d\vec{r} \,d\vec{r}_1 \nonumber \\
 && + \; {Z^\prime \over 2}\sum_{i,j,s}\iint  \dfrac{\phi^\star_{i\,s}
(\vec{r})\,\phi^\star_{j\,s}(\vec{r}_1)\,\phi_{i\,s}
(\vec{r}_1)\,\phi_{j\,s}(\vec{r})} {|\vec{r}-\vec{r}_1|}\,
d\vec{r} \,d\vec{r}_1\nonumber\\
&& - \; {B^\prime \over 2}\sum_{i,j,s} \iint {\phi^\star_{i\,s}(\vec{r})
\,\phi^\star_{j\,s}(\vec{r}_1)\,\phi_{i\,s}(\vec{r}_1)\,\phi_{j\,s}
(\vec{r})}\, {|\vec{r}-\vec{r}_1|} \,d\vec{r} \,d\vec{r}_1
\label{har_fock_ham}
\end{eqnarray}
where, we have used
\begin{eqnarray*}
\langle T \rangle &=& \sum_{i,s} \left\langle \phi_{i\,s}(\vec{r})
\left\vert -{1\over 2m_{u/d}} \nabla^2 \right\vert \phi_{i\,s}(\vec{r})
\right \rangle \\
\rho(\vec{r}) &=& \sum_{i,s} \left\vert \phi_{i\,s}(\vec{r}) \right
\vert^2, \;\;\;\;
V_{\textmd{ext}}(\vec{r}) = -{2\alpha\over 3r}+ {\beta r \over 2} \\
Z^\prime &=& {2\alpha^\prime\over 3} \;\;\textmd{and}\;\;\;
B^\prime = {\beta^\prime \over 2}.
\end{eqnarray*}
In contrast to the helium atom, the presence of linear $r$-terms
in the Hamiltonian leads to additional exchange-energy terms in
the calculation. With these linear $r$-terms in, the Hartree-Fock
equation becomes
\begin{eqnarray}
E\,\phi_{i\,s}(\vec{r}) &=& \left[ -{1\over 2m_{u/d}}\nabla^2 +
V_{\textmd{ext}}(\vec{r}) -Z^\prime \int {\rho(\vec{r}_1)\over |\vec{r}-
\vec{r}_1|} \, d\vec{r}_1 + B^\prime \int \rho(\vec{r}_1) \,|\vec{r}-
\vec{r}_1| \,d\vec{r}_1 \right]\, \phi_{i\,s}(\vec{r}) \nonumber\\
 && - \; B^\prime \sum_{j,s} \int {\phi^\star_{j\,s}(\vec{r}_1)\,
\phi_{i\,s}(\vec{r}_1)\, \phi_{j\,s}(\vec{r})}\, {|\vec{r}-\vec{r}_1|}
\, d\vec{r}_1 \nonumber\\
 && + \; Z^\prime\sum_{j,s}\int \dfrac{\phi^\star_{j\,s}(\vec{r}_1)
\,\phi_{i\,s}(\vec{r}_1)\, \phi_{j\,s}(\vec{r})} {|\vec{r}-\vec{r}_1|}
\,d\vec{r}_1 \label{mod_har_fock_ham}
\end{eqnarray}
We solve for $E$ in equation (\ref{mod_har_fock_ham}) iteratively
and, eventually, the $\Lambda_b$ mass is calculated from
\begin{equation}
M_{\Lambda_b} = m_b + 2m_{u/d}\,+ E \label{mass_lambda_b} 
\end{equation}
The PDG value of $\Lambda_b (5620)$ is obtained by setting the $m_{u/d}$
to 0.157 GeV.
\begin{table}[h]
\caption{Comparison of $m_{u/d}$ obtained from various lattices with
quark mass parameters in the equations (\ref{schro_eq}) and
(\ref{mod_har_fock_ham}).}
\begin{ruledtabular}
\begin{tabular}{ccccc}
\multirow{2}{*}{Lattice} & \multicolumn{2}{c}{$B$ meson: $m_{u/d} =
227$ MeV} & \multicolumn{2}{c}{$\Lambda_b$ baryon: $m_{u/d} = 157$ MeV} \\
 & $am_{u/d}$ & $m_{u/d}$ (MeV) & $am_{u/d}$ & $m_{u/d}$ (MeV) \\ \hline
$16^3 \times 48$ & 0.155 & 204 & 0.105 & {138} \\
$20^3 \times 64$ & 0.118 & 194 & 0.083 & 137 \\
$28^3 \times 96$ & 0.087 & 191 & 0.064 & 143 \\
\end{tabular}
\end{ruledtabular}
\label{tab:mq_compare}
\end{table}
In Table \ref{tab:mq_compare}, we compare the nonperturbatively tuned
$m_{u/d}$ on our lattices with those obtained by solving the equations
(\ref{schro_eq}) and (\ref{mod_har_fock_ham}). The bare lattice light
quark masses cannot be directly compared to the parameter $m_{u/d}$
in these equations mainly because of the use of renormalized $b$
quark mass (in $\overline{\text{MS}}$ scheme) in the Hartree-Fock
calculation. Therefore, the $m_{u/d}$'s in the above calculation
return a sort of ``renormalized constituent" quark mass. Nonetheless
it is obvious that we need two different $m_{u/d}$ for two different
systems, namely $B$ and $\Lambda_b$. So by comparing the two sets,
we simply wish to point out that the lattice tuned $m_{u/d}$'s are in
same order of magnitude as Schr\"{o}dinger equation based quark model
but have a difference of 10 -- 15\%. This helps us to understand the
possible physics behind two different tunings of light quark mass in
determining the masses of single bottom hadron(s) and double bottom
tetraquark.

\subsection{$bb\bar{u}\bar{d}$ spectrum} \label{subsec_spectra}
\begin{center}
\begin{figure}[h]
\includegraphics[scale=0.9]{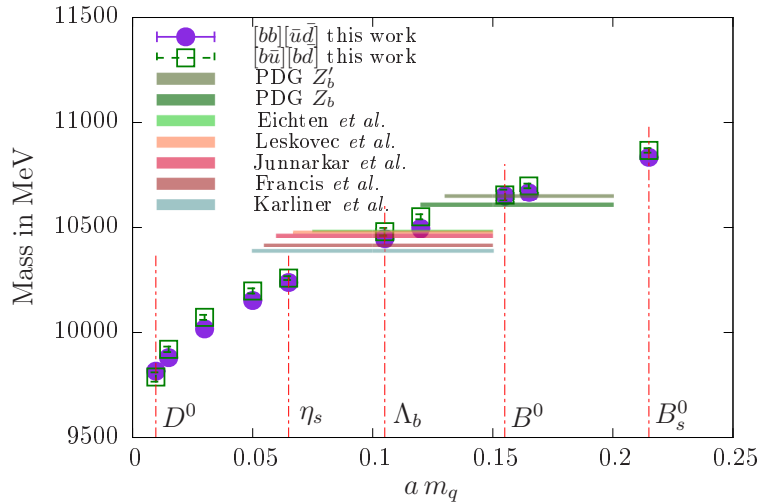}
\caption{Variation of $bb\bar{u}\bar{d}$ mass at various $am_{u/d}$
in $16^3 \times 48$ lattice. $\Lambda_b$-tuned tetraquark states almost
overlap with many of the quark model and lattice calculations, namely
Eichten {\it et al.} \cite{eichten}, Leskovec {\it at al.} \cite{luka},
Junnarkar {\it et al.} \cite{junnarkar}, Francis {\it et al.}
\cite{francis1,francis2}, Karliner {\it at al.} \cite{karliner}. The
$B$-tuned states instead coincide with $Z_b,\,Z_b^\prime$ PDG results
\cite{pdg}.}
\label{fig:tune}
\end{figure}
\end{center}

A plot of variation of $bb\bar{u}\bar{d}$ mass with various $am_{u/d}$,
including the $\Lambda_b$ and $B$ tuned values is shown in Fig.
\ref{fig:tune}. Here we make a naive comparison of our data with the
earlier quark model, lattice calculations and the PDG values, and it
shows an interesting trend.

Firstly, PDG $Z_b,\,Z_b^\prime$ and the lattice states,
consisting of $[bb] /[\bar{b}\bar{b}]$ heavy tetraquark
systems, are clustered around two different masses. Our data at $B$-meson
tuning point coincides with the PDG $Z_b\,(10610)$ and $Z_b^\prime$ (10650)
states aligning with the idea that they decay mostly into $\bar{B}B^*$ and
$\bar{B}^*B^*$ respectively, possibly indicating molecular nature of
the state. However, our tetraquark state with $\Lambda_b$ tuning
overlaps mostly with other lattice results indicating the possibility
of capturing a bound tetraquark state $[bb][\bar{l}_1\bar{l}_2]$ much
like the $b[{l}{l}]$ state of $\Lambda_b$. The effective masses of the
states, obtained from $\mathcal{O}_D$ and $\mathcal{O}_{M_1}$, when
compared with the $B-B^*$ threshold, we find $|D\rangle$ to exhibit 
a {\it shallow} bound state while $|M_1\rangle$ is just marginally above. 
The majority of the lattice results \cite{francis1,francis2,junnarkar,luka} 
are found to be below this threshold as is obvious from the 
Fig. \ref{fig:tune}.

\begin{center}
\begin{figure}[H]
\includegraphics[scale=1.0]{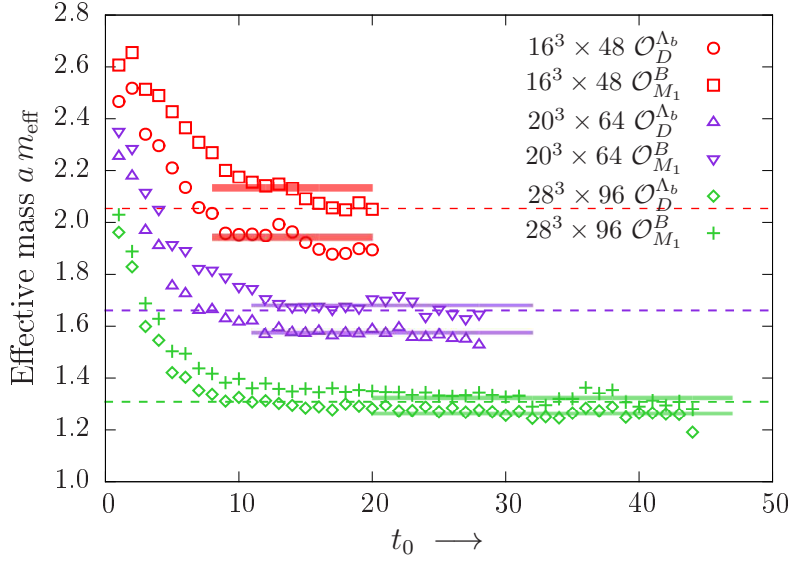}
\caption{Effective mass plot of the states of the operators
$\mathcal{O}_D$ and $\mathcal{O}_{M_1}$ calculated on $16^3\times 48$,
$20^3 \times 64$ and $28^3\times 96$ lattices. Dashed lines are
$B-B^*$ thresholds for different lattices, see
Table \ref{tab:ttq_mass}.
For easy viewing, the effective masses and thresholds on $20^3\times 64$
(purple colored) are multiplied by a common factor of 0.85, while that
of $28^3\times 96$ (green colored) by 0.70.}
\label{fig:efm}
\end{figure}
\end{center}

To this end, in Fig. \ref{fig:efm} we plot the effective masses of these
two states obtained at different lattice spacings. The colored bands
represent fitted $am_{\mathrm{eff}}$ values. The superscripts $\Lambda_b$
and $B$ denote the light quark tuning. The dotted lines represent the
lattice thresholds. The lattice threshold is defined
as $M_{B} + M_{B^*}$ of the non-interacting $B-B^*$ system and consequently
constructed entirely from $B$-meson tuned $am_{u/d}$ and $\Upsilon -
\eta_b$ tuned $m_b$. We want to mention here in passing that on two
occasions  we used $\Lambda_b$ tuning for operators $\mathcal{O}_{M_1/
M_2}$ -- ($i$) to choose $t_0$ for GEVP analysis and ($ii$) to determine
relative contribution to the bound tetraquark ground state.

In the Table \ref{tab:ttq_mass}, we present our results of the tetraquark
states corresponding to the operators given in the expressions (\ref{bsb},
\ref{bsbs}, \ref{Qpi}). We call these states
$\mathcal{O}_X^\dagger | \Omega \rangle \equiv |X\rangle$ as trial
states, which will later be subjected to variational analysis. The
$|\Omega \rangle$ is the vacuum state. We use two-exponential
uncorrelated fit to the correlation functions, the fitting range being
chosen by looking at the positions of what we consider plateau in the
effective mass plots. In the columns showing various lattices, we present
the masses both in lattice unit $a\,E_{\textmd{latt}}$ and physical unit
$M_{\textmd{latt}}$ in MeV, the notations being introduced in equation
(\ref{mass_formula}). The errors quoted are statistical, calculated
assuming the lattice configurations of different lattice spacings
are statistically uncorrelated. The second column shows the tuning
used for the corresponding states. In the last column we provide
the masses averaged over all the lattice ensembles.
\begin{table}[h]
\caption{Masses of tetraquark states for different $am_{u/d}$ tuning in
lattice unit $aE_{\textmd{latt}}$ and $M_{\textmd{latt}}$
in MeV. We also include the $B$ and $B^*$ states that are
used for threshold calculation.}
\begin{ruledtabular}
\begin{tabular}{lcccccccc}
\multirow{2}{*}{Operators} & \multirow{2}{*}{Tuning} &
\multicolumn{2}{c}{$16^3\times 48$} &
\multicolumn{2}{c}{$20^3\times 64$} & \multicolumn{2}{c}{$28^3\times
96$} & Average \\
 & & $aE_{\textmd{latt}}$ & $M_{\textmd{latt}}$ &
 $aE_{\textmd{latt}}$ & $M_{\textmd{latt}}$ &
 $aE_{\textmd{latt}}$ & $M_{\textmd{latt}}$ & (MeV) \\ \hline
$\mathcal{O}_D = [bb][\bar{u}\bar{d}]$ & $\Lambda_b$ & 1.944(5)
 & 10418(7) & 1.852(3) & 10422(5) & 1.803(5) & 10407(11) & 10417(9) \\
$\mathcal{O}_{M_1} = [b\bar{u}][b\bar{d}]$ & $B$ & 2.133(7) & 10667(10)
 & 1.977(4) & 10628(5) & 1.892(6) & 10602(13) & 10638(27) \\
$\mathcal{O}_{M_2} = \epsilon_{ijk}[b\bar{u}]_j[b\bar{d}]_k$ & $B$ &
 2.124(7) & 10655(8) & 1.974(4) & 10623(5) & 1.890(5) & 10560(10)
& 10623(35) \\
$\mathcal{O}_B = b\,\gamma_5\,\bar{u}$ & $B$ & 1.022(3) & 5274(4) &
0.974(3) & 5290(3) & 0.931(3) & 5268(3) & 5279(10) \\
$\mathcal{O}_{B^*} = b\,\gamma_k\,\bar{u}$ & $B^*$ & 1.032(3) & 5288(4)
& 0.980(3) & 5300(4) & 0.938(2) & 5284(3) & 5292(8)\\ \hline
$M_B + M_{B^*}$ & & 2.054(3) &
10562(4) & 1.954(3) & 10590(5) & 1.869(3) & 10552(4) & \\
\end{tabular}
\end{ruledtabular}
\label{tab:ttq_mass}
\end{table}

From the Fig. \ref{fig:efm} and Table \ref{tab:ttq_mass} it is clear
that the trial state generated by our $\mathcal{O}_D$ operator is below
$B-B^*$ threshold which possibly indicates a bound state. On the other
hand, the states for $\mathcal{O}_{M_1}$ and $\mathcal{O}_{M_2}$ are just
above it. We tabulate the difference of the masses from their
respective thresholds $\Delta M_{D/M_1/M_2} = M_{D/M_1/M_2} - M_B -
M_{B^*}$ in the Table \ref{tab:efm_dE}. In this table, we calculated
the following correlator ratio to determine the mass differences which
gives us an estimate of the binding energy \cite{beane},
\begin{equation}
C_{X-B-B^*}(t) = \frac{C_X(t)}{C_B(t) \times C_{B^*}(t)} \sim
e^{-(M_X - M_B - M_{B^*})\,t} \label{corr_ratio}
\end{equation}
It has been observed \cite{iritani} that the expression
(\ref{corr_ratio}) used to determine $\Delta M_X$ can possibly lead to
false plateaus because of $B-B^\star$ scattering states contributing
differently in $|D\rangle,\,|M_1\rangle,\,|M_2\rangle$ excited
states which might persist at large $t$. In the present analysis, we
have assumed these contributions are of same order of magnitude and
cancel each other at moderately large $t$.
\begin{table}[h]
\caption{Mass differences of ``bound" $|D\rangle$ and ``molecular"
$|M_1\rangle, \,|M_2\rangle$ trial states from $B-B^*$ threshold.
The subscript $X$ denotes any of the $D,\,M_1,
\,M_2$. The $\Delta M_X$ are calculated from the masses and
threshold given in Table \ref{tab:ttq_mass}.}
\begin{ruledtabular}
\begin{tabular}{lcccl}
Operators & Lattices & $a\,\Delta M_X$ & $\Delta M_X$ (MeV) &
 $\overline{\Delta M_X}$ (MeV) \\
\hline
\multirow{3}{*}{$\mathcal{O}_D$} & $16^3\times 48$ & $-0.125(12)$ &
 $-164(16)$ & $-167(19)$ this work \\
 & $20^3\times 64$ & $-0.108(10)$ & $-177(16)$ & $-215(12)$ \cite{karliner} \\
 & $28^3\times 96$ & $-0.070(10)$ & $-155(22)$ & $-189(10)$ \cite{francis1} \\
 & & & & $-143(34)$ \cite{junnarkar} \\
 & & & & $-128(34)$ \cite{luka} \\ \hline
\multirow{3}{*}{$\mathcal{O}_{M_1}$} & $16^3\times 48$ & $0.070(12)$ &
 92(16) & 65(29) this work \\
 & $20^3\times 64$ & $0.026(11)$ & 43(18) & see Table VI \cite{luka}\\
 & $28^3\times 96$ & $0.024(9)$ & 53(20) & \\ \hline
\multirow{3}{*}{$\mathcal{O}_{M_2}$} & $16^3\times 48$ & $0.070(16)$ &
 92(21) & 63(30) this work \\
 & $20^3\times 64$ & $0.022(9)$ & 36(20) & \\
 & $28^3\times 96$ & $0.020(10)$ & 44(21) & \\
\end{tabular}
\end{ruledtabular}
\label{tab:efm_dE}
\end{table}

\noindent
In the last column of
Table \ref{tab:efm_dE}, we calculate our lattice average of $\Delta
M_X$ in MeV and compare with some of the previous lattice results.
To our knowledge, the binding energies of the $|M_1\rangle$ and
$|M_2\rangle$ states have been calculated in the framework of chiral
quark model \cite{yang} for $B-\bar{B}^*$ and $B^*-\bar{B}^*$ states
but there are no lattice results. The binding energies for the first
excited states, along with the ground states, obtained on different
lattice ensembles are given in \cite{luka}. Though their tuning of
light quark mass is very different compared to ours, still we can use
their result as a reference.

Our binding energy for the bound tetraquark state $|D \rangle \,
([bb][\bar{u}\bar{d}])$ lies somewhere in the middle of the
previously quoted lattice results. The statistical errors of the
molecular states $|M_1\rangle$ and $|M_2\rangle$ are rather large
but still they tentatively indicate non-bound molecular nature of
the states. We will revisit the binding energy calculation for the
molecular state(s) after variational analysis of the $\mathcal{O}_{M_1}
\times \mathcal{O}_{M_2}$ correlation matrix.

As we know, on lattice the states having the same quantum numbers
can mix and, therefore, a GEVP analysis can help resolve the issue
of mutual overlap of various states on the energy eigenstates. In
this work, rather than the energies of the eigenstates, we are more
interested to learn the overlap of our trial states, namely $|D\rangle,
\,|M_1\rangle$ and $|M_2\rangle$ on the first few energy eigenstates,
where $|0\rangle$ is the ground and $|1\rangle,\,|2\rangle$
etc. are the excited states.

\subsection{Variational analysis} \label{subsec_gevp}
For the 2-bottom tetraquark system with quantum number $1^+$, we
consider the three local operators -- ``good" diquark $\mathcal{O}_{D}$,
molecular $\mathcal{O}_{M_1}$ and vector meson kind $\mathcal{O}_{M_2}$
as defined above in the expressions (\ref{bsb} -- \ref{Qpi}) -- to
capture the ground state $(\vert 0 \rangle,\, \mathcal{E}_0)$ and
possibly the first excited state $(\vert 1 \rangle,\,\mathcal{E}_1)$.

As is generally understood, these operators are expected to have overlap
with the desired ground and excited states of the tetraquark system of
our interest. The variational analysis can be performed to determine
the eigenvalues and the eigenvectors from the states formed
by lattice operators. This is typically achieved by constructing a
correlation matrix involving the lattice operators $\mathcal{O}_X$ and
$\mathcal{O}_Y$,
\begin{equation}
C_{XY}(t) = \left\langle \mathcal{O}_X(t) \,\mathcal{O}_Y^\dagger(0)
\right \rangle 
 = \sum_{n=0}^\infty \langle \Omega \vert \mathcal{O}_X \vert n
\rangle \langle n \vert \mathcal{O}_Y^\dagger \vert \Omega \rangle
\,e^{-E_nt} \label{correxp}
\end{equation}
where $X,Y$ can be any two combinations of $D,\,M_1,\,
M_2$ in the expressions (\ref{bsb} -- \ref{Qpi}). The terms
$\langle n \vert \mathcal{O}_X^\dagger \vert \Omega\rangle$ are the
coefficients of expansion of the trial states $\mathcal{O}_X^\dagger
\vert \Omega \rangle$, written in terms of the energy eigenstates
$\vert n \rangle$ as,
\begin{equation}
\mathcal{O}_X^\dagger \vert \Omega \rangle = \sum_n \vert n \rangle
\langle n\vert \mathcal{O}_X^\dagger \vert \Omega \rangle \equiv
\sum_n Z_X^n \vert n \rangle \label{trialbasis}
\end{equation}
Presently, we are interested in expressing the energy eigenstates
in terms of the trial states to understand the contribution of each
to the former. If we confine ourselves to the first few energy
eigenstates, we can write
\begin{equation}
\vert m \rangle = \sum_X v_m^X \,\mathcal{O}_X^\dagger \vert \Omega
\rangle \;\;\Rightarrow \;\; \langle l \vert m \rangle = \delta_{lm}
\approx \sum_X v_m^X \,Z_X^l \label{energybasis}
\end{equation}
The $v_m^X$ are equivalent to the eigenvector components obtained by
solving GEVP w.r.t a suitably chosen reference time $t_0$
\cite{alexandrou},
\begin{equation}
C(t)\,v_m(t,t_0) = \lambda_m(t)\,C(t_0)\,v_m(t,t_0). \label{gevp}
\end{equation}
The eigenvalues $\lambda_m(t)$ are directly related to the energy of
the $m$-th state, {\it i.e.} ground and the first few excited states,
of our system through the relation
\begin{equation}
\lambda_m(t) = A_m\,e^{-\mathcal{E}_m (t-t_0)} \label{eval_e}
\end{equation}
The component of eigenvectors $v_m(t,t_0)$ gives information about the
relative overlap of the three local operators to the $m$-th
eigenstate. The eigenvectors $v_m$'s are normalized to 1.
\begin{center}
\begin{figure}[H]
\includegraphics[scale=1.0]{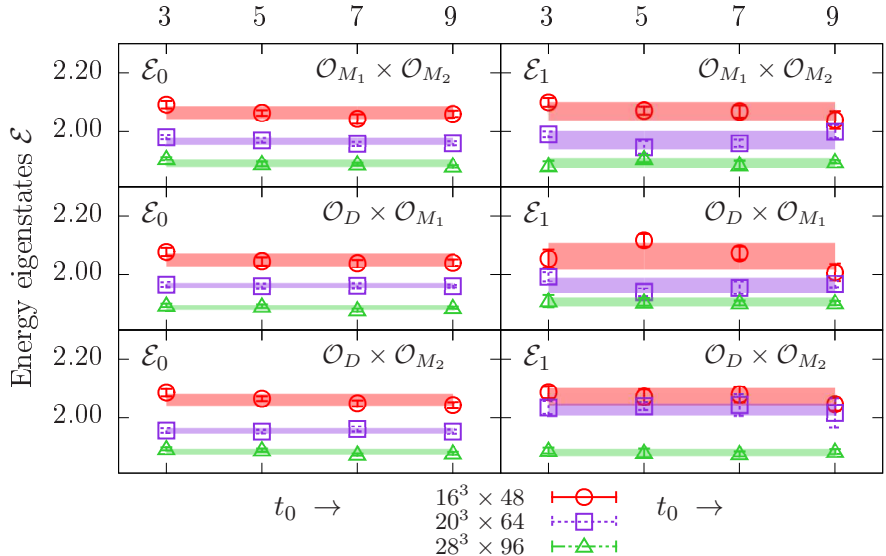}
\caption{Variation of ground and excited state energies $\mathcal{E}_m$
of the equation (\ref{eval_e}) with $t_0$, obtained by solving $2 \times 2$
GEVP of the correlation matrices $\mathcal{O}_{M_1}\times \mathcal{O}_{M_2}$,
$\mathcal{O}_{D} \times \mathcal{O}_{M_1}$ and $\mathcal{O}_{D} \times
\mathcal{O}_{M_2}$. In this plot we used $\Lambda_b$-tuned $am_{u/d}$
for all the operators.}
\label{fig:t0}
\end{figure}
\end{center}

To determine the parameter $t_0$, we solve the GEVP and found that the
ground and excited state energies are almost independent for $t_0 = 3,
5,7,9$ as demonstrated in the Fig. \ref{fig:t0}. In this plot, we showed
our results for $\Lambda_b$-tuned $am_{u/d}$ for all the operators
$\mathcal{O}_X$. As discussed before, we have used the
$\Lambda_b$ tuning whenever all three $\mathcal{O}_D,\,\mathcal{O}_{
M_1},\,\mathcal{O}_{M_2}$ operators are
made use of. We chose $t_0 = 5$ for our calculations. To cross-check
our choice of $t_0$, we also have $B$ tuned runs and found it to be
consistent.

The GEVP analysis has been carried out in two steps because of differences
in the tuning of $am_{u/d}$ for the ``molecular" states $|M_1\rangle,
\,|M_2\rangle$ and ``good" diquark state $|D\rangle$. In
the first step, we will do a GEVP with the $B$-tuned molecular operators
and determine the difference of its lowest energy state from the
threshold, as these states are found to coincide with experimentally
observed states. In the next step, we have done the GEVP analysis with
all three operators using $\Lambda_b$ tuning to understand the state(s)
below the threshold.

\begin{table}[h]
\caption{Energy eigenvalues in lattice unit from GEVP
analysis of the
$B$-tuned $\mathcal{O}_{M_1}\times \mathcal{O}_{M_2}$ and $\Lambda_b$-tuned
$\mathcal{O}_D \times \mathcal{O}_{M_1}\times \mathcal{O}_{M_2}$.}
\begin{ruledtabular}
\begin{tabular}{cccccc}
Correlation matrix & Tuning & Energy & $16^3\times 48$ & $20^3\times 64$
 & $28^3\times 96$ \\ \hline
\multirow{2}{*}{$\mathcal{O}_{M_1} \times \mathcal{O}_{M_2}$} &
\multirow{2}{*}{$B$ meson} & $\mathcal{E}_0$ & 2.063(10) & 1.959(12) &
 1.888(7) \\
 & & $\mathcal{E}_1$ & 2.071(10) & 1.969(20) & 1.906(18) \\ \hline
\multirow{3}{*}{$\mathcal{O}_D \times \mathcal{O}_{M_1}\times
\mathcal{O}_{M_2}$} & \multirow{3}{*}{$\Lambda_b$ baryon} &
$\mathcal{E}_0$ & 1.898(7) & 1.846(5) & 1.784(12) \\
 & & $\mathcal{E}_1$ & 1.905(10) & 1.851(7) & 1.816(8) \\
 & & $\mathcal{E}_2$ & 1.917(18) & 1.856(15) & 1.820(22) \\
\end{tabular}
\end{ruledtabular}
\label{tab:dm1m2gevp}
\end{table}

In the Table \ref{tab:dm1m2gevp} we have shown our GEVP
results of $B$-tuned $\mathcal{O}_{M_1} \times \mathcal{O}_{M_2}$ and the
$\Lambda_b$-tuned $\mathcal{O}_D \times \mathcal{O}_{M_1} \times
\mathcal{O}_{M_2}$ correlation matrices. The energy eigenstates
$\mathcal{E}_{0,1,2}$ correspond to the $\mathcal{E}_m$ of the
expression (\ref{eval_e}). In the Table \ref{tab:dm1m2gevp_dE}, we
calculated the energy difference of the eigenstates $|0\rangle ,\,
|1\rangle$ etc. from their corresponding thresholds. We often find
the energies of the highest states are very noisy and consequently
the seperation from the thresholds $\Delta \mathcal{E}$ have large
errors, hence their entries are kept vacant. We can only reliably
quote the lowest for $2 \times 2$, and first two lowest energies
for $3 \times 3$ correlator matrices. 

\begin{table}[h]
\caption{Energy differences from the $B-B^*$ threshold
of the GEVP values of Table \ref{tab:dm1m2gevp} for the
$\mathcal{O}_{M_1}\times \mathcal{O}_{M_2}$ and $\Lambda_b$-tuned
$\mathcal{O}_D \times \mathcal{O}_{M_1}\times \mathcal{O}_{M_2}$
correlation matrices. Threshold values are taken from Table
\ref{tab:ttq_mass}.}
\begin{ruledtabular}
\begin{tabular}{cccccccc}
Lattice & Unit & Threshold & \multicolumn{2}{c}{$\mathcal{O}_{M_1} \times
\mathcal{O}_{M_2}$} & \multicolumn{3}{c}{$\mathcal{O}_D \times
\mathcal{O}_{M_1} \times \mathcal{O}_{M_2}$} \\ \cline{4-8}
 & & & $\Delta \mathcal{E}_0$ & $\Delta \mathcal{E}_1$ & $\Delta
\mathcal{E}_0$ & $\Delta \mathcal{E}_1$ & $\Delta \mathcal{E}_2$ \\ \hline
$16^3 \times 48$ & lattice & 2.054(3) & 0.010(7) & 0.016(6) & $-0.154(10)$
& $-0.149(15)$ & $-0.137(23)$ \\
(0.15 fm) & MeV & 10562 & 13(9) & 21(8) & $-202(13)$ & $-196(20)$
& -- \\ \hline
$20^3 \times 64$ & lattice & 1.954(3) & 0.010(9) & -- & $-0.110(9)$
& $-0.104(10)$ & $-0.099(19)$ \\
(0.12 fm) & MeV & 10590 & 16(15) & -- & $-181(15)$ & $-173(17)$ & -- \\ \hline
$28^3 \times 96$ & lattice & 1.869(3) & 0.012(10) & -- & $-0.085(10)$
& $-0.053(10)$ & -- \\
(0.09 fm) & MeV & 10552 & 26(22) & -- & $-186(22)$ & $-117(22)$ & -- \\
\end{tabular}
\end{ruledtabular}
\label{tab:dm1m2gevp_dE}
\end{table}

\begin{center}
\begin{figure}[h]
\includegraphics[scale=0.9]{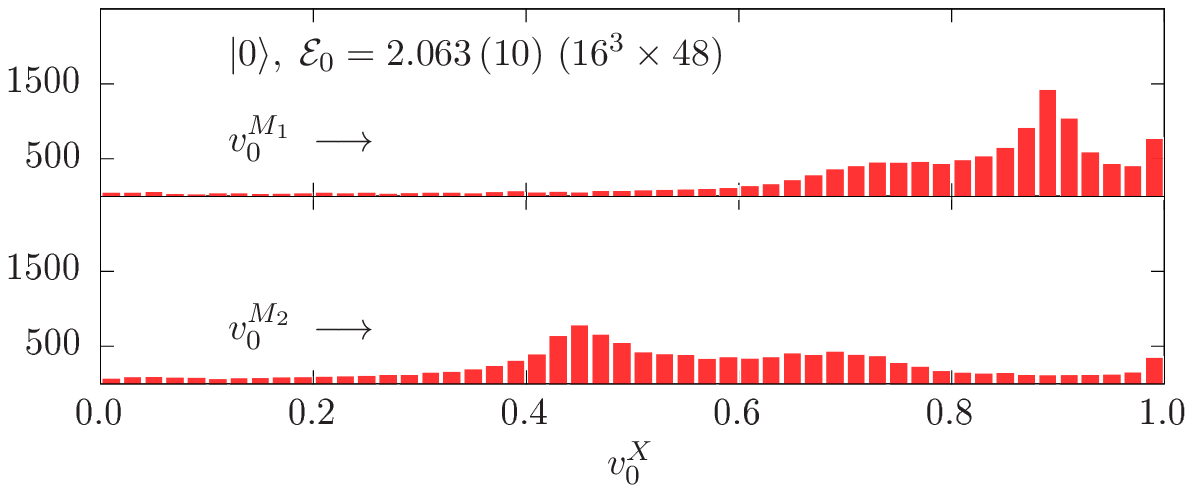} \\
\vskip -0.9in
\includegraphics[scale=0.9]{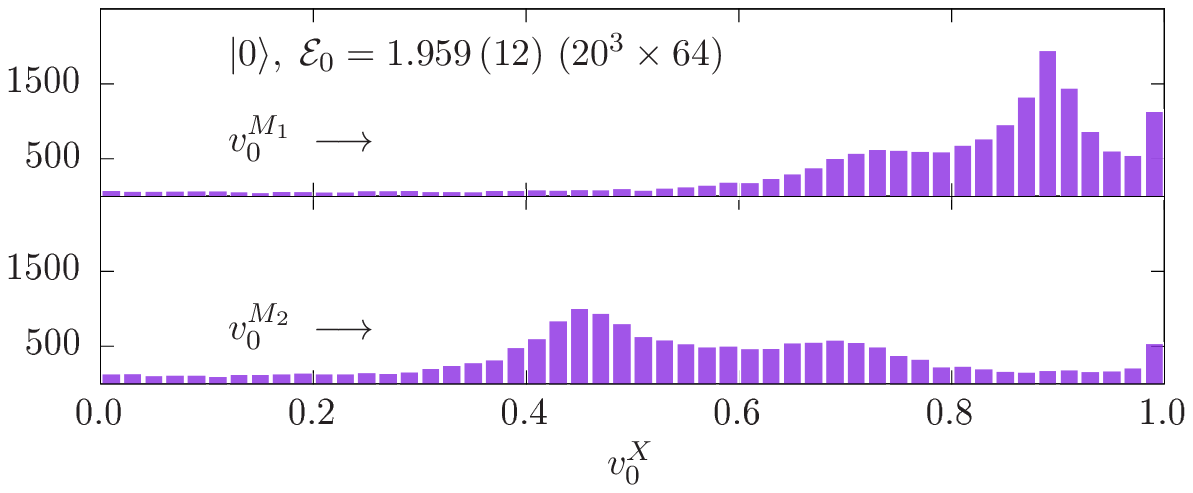} \\
\vskip -0.9in
\includegraphics[scale=0.9]{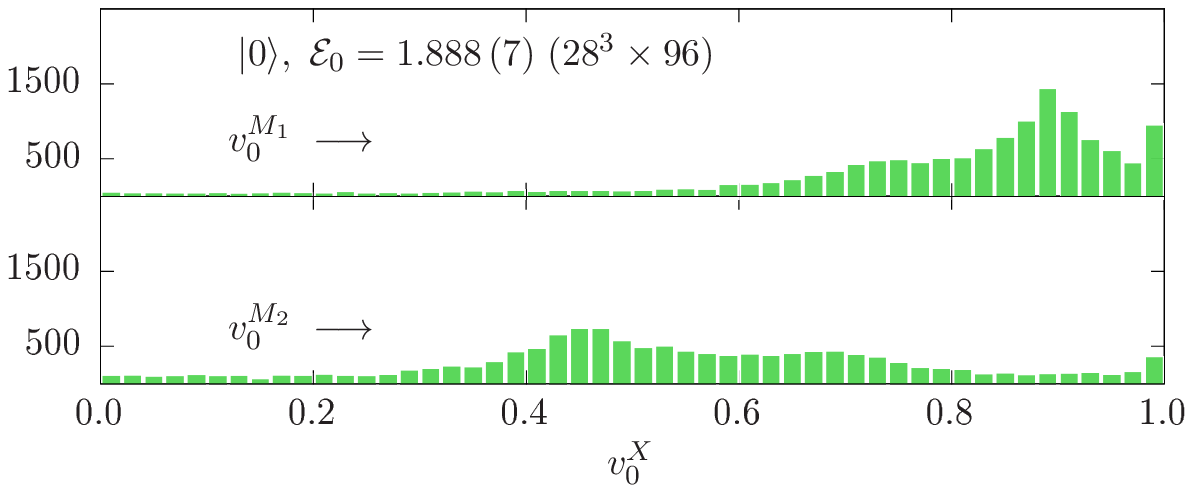}
\vskip -0.9in
\caption{The histogram plots of the normalized components $(v_0^{M_1},
\,v_0^{M_2})$ which define the energy eigenstate $|0\rangle = v_0^{M_1}
|M_1\rangle + v_0^{M_2} |M_2\rangle$.}
\label{fig:m1m2ev0}
\end{figure}
\end{center}

Next we look at the contribution of $\mathcal{O}_{M_1}$ and
$\mathcal{O}_{M_2}$ in constructing the lowest molecular energy
eigenstate $\vert 0\rangle$. In the Fig. \ref{fig:m1m2ev0}, we plot
the histogram of the components of normalized eigenvectors
$v_0 = (v_0^{M_1},\, v_0^{M_2})$ corresponding to the lowest energy
$\mathcal{E}_0$ for all three lattices. Assuming that the coefficients
$v_0^{M_1,\,M_2}$ approximately remain the same on all time slices
and for all the individual gauge configurations of an ensemble, the
histogram figures are obtained by
plotting the $M_1,\,M_2$ components of normalized eigenvector
$v_0$ for all time points and individual gauge configurations.
As is expected, all three lattices return identical histogram of the
coefficients and hence, in the subsequent histogram plots we will show
only the results from $28^3 \times 96$. The eigenvector component
$v_0^{M_1}$ shows a peak around 0.9 indicating the lowest energy state
$\vert 0 \rangle$ receives dominant contribution from $|M_1\rangle$
trial state.
We recall here that $\mathcal{O}_{M_1}$ corresponds to the $B-B^*$
molecular state as defined in the equation (\ref{bsb}).

\begin{center}
\begin{figure}[h]
\includegraphics[scale=0.9]{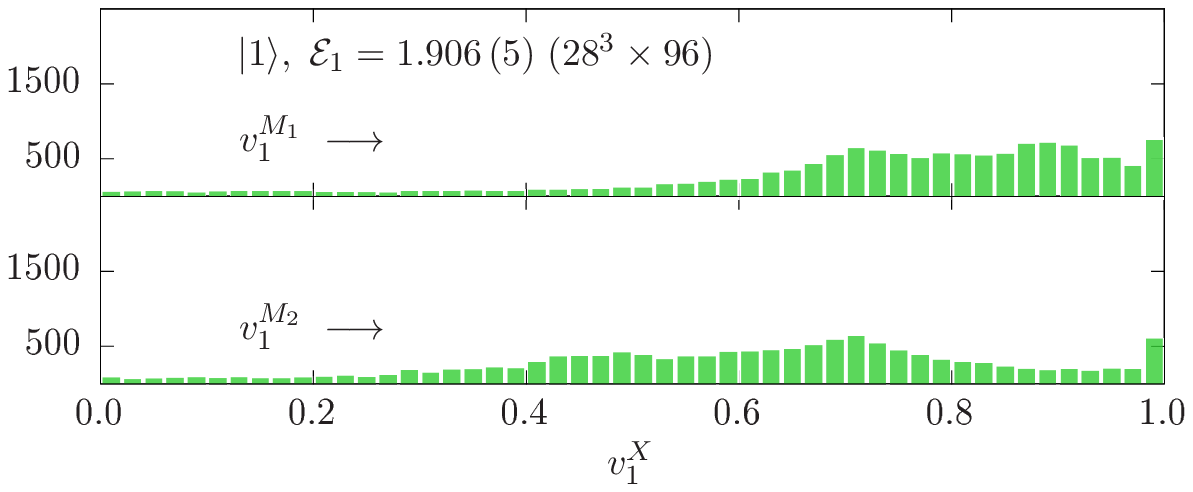}
\vskip -0.6in
\caption{The histogram plot of $v_1^{M_1}$ and $v_1^{M_2}$ that define
the energy eigenstate $|1\rangle = v_1^{M_1} |M_1\rangle + v_1^{M_2}
|M_2\rangle$.}
\label{fig:m1m2ev1}
\end{figure}
\end{center}

However, the first excitation $|1\rangle$, for which our data is
rather noisy to reliably estimate $\Delta \mathcal{E}$, the $|M_1\rangle$
and $|M_2\rangle$ states appear to have comparable contribution
and are broadly distributed over different time slices and vary
significantly over configurations. This is evident from the histogram
plot in the Fig. \ref{fig:m1m2ev1}. This may have a bearing
with the fact that above the threshold, the $Z_b$ tetraquark can
couple to multiple decay channels resulting in a broad spectrum.

Including $\mathcal{O}_D$ along with the $\mathcal{O}_{M_1}$ and
$\mathcal{O}_{M_2}$ to form a $3 \times 3$ correlation matrix requires
using $\Lambda_b$ tuned $am_{u/d}$ in all three trial states.
When we are exploring pure molecular states, we have
used just $\mathcal{O}_{M_1} \times \mathcal{O}_{M_2}$ correlation
matrix with $B$ tuning. But for the bound state(s) the $\mathcal{O}_{M_1}$
and $\mathcal{O}_{M_2}$ operators are likely to have contributions to
the bound ground state along with the $\mathcal{O}_D$. Certainly,
a $B$-tuned bound $|D\rangle$ state
above threshold is not well-defined and we find it has statistically
small and varying overlap with the energy eigenstates much like in
Fig. \ref{fig:m1m2ev1}. On the other hand, $\Lambda_b$ tuned molecular
states can possibly have finite overlap to the states below the
threshold. However, we always expect dominance of $|D\rangle$ in $\vert 0
\rangle$ because of the difference in construction of wave functions
of the $|M_1\rangle$ and $|M_2\rangle$.
\begin{center}
\begin{figure}[h]
\includegraphics[scale=0.8]{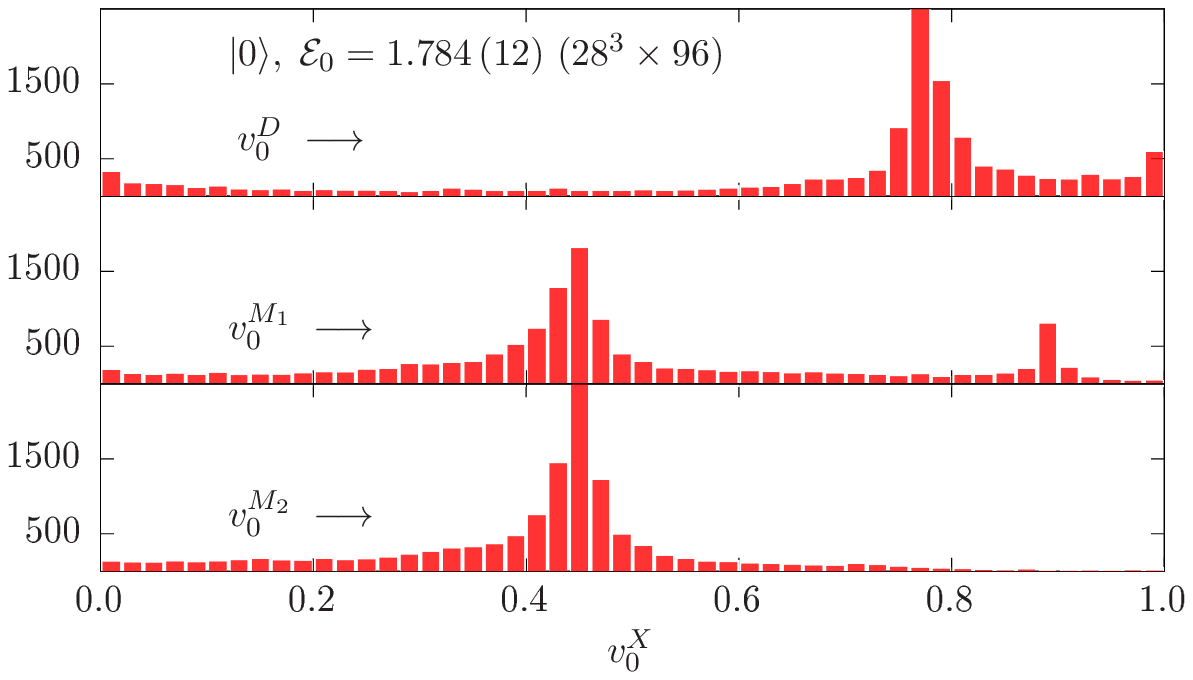} \\
\vskip -0.2in
\includegraphics[scale=0.8]{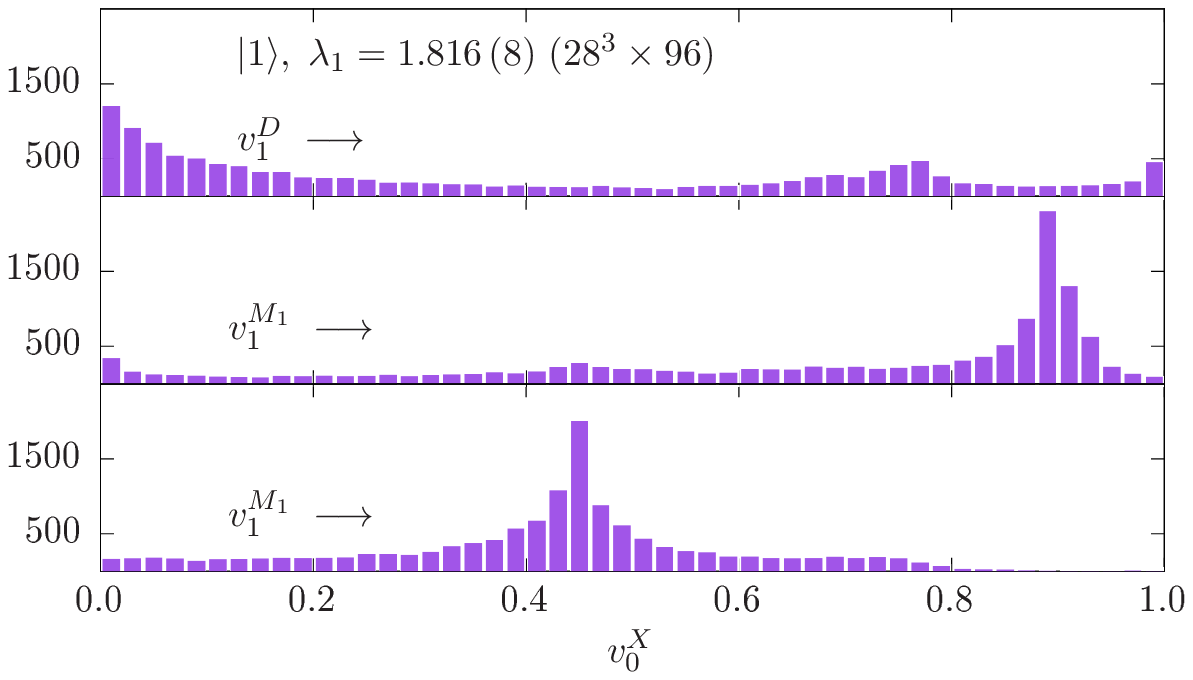} \\
\vskip -0.2in
\includegraphics[scale=0.8]{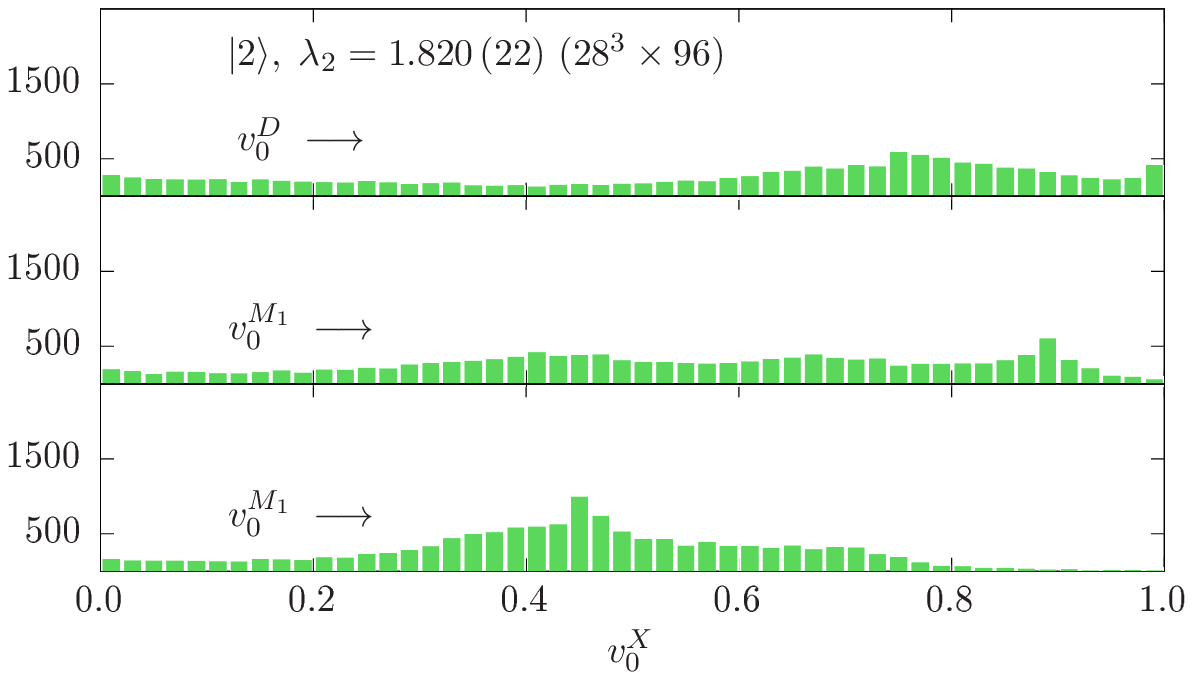} \\
\vskip -0.2in
\caption{Histogram plots of the normalized eigenvector components $v_0^X$,
$v_1^X$ and $v_2^X$ of $3 \times 3$ correlation matrix, where $X = D,\,
M_1,\,M_2$, on $28^3 \times 96$ lattice.}
\label{fig:dm1m2ev}
\end{figure}
\end{center}
The histogram of the eigenvector components of $3 \times 3$ correlation
matrix are shown in the Fig. \ref{fig:dm1m2ev} for the $28^3 \times
96$ lattices. The lowest energy eigenstate $\vert 0 \rangle$ is clearly
dominated by $|D\rangle$ showing peak around 0.8, although it receives
sizeable overlap from both $|M_1\rangle$ and $|M_2\rangle$ peaking
around 0.45. But overlap of $|D\rangle$ on $\vert 1 \rangle$ is rather
small and it is mostly molecular $|M_1\rangle$ despite the excited
state energy $\mathcal{E}_1$ is below the threshold. Our
data for $\vert 2 \rangle$ is too noisy to extract much information.
Based on this $\Lambda_b$ tuned $3 \times 3$ GEVP analysis, the binding
energy for the ground state obtained is $-189\,(18)$ MeV.

\section{Summary} \label{sec_summary}
In this work we have attempted to study two possible $bb\bar{u}\bar{d}$
tetraquark states -- one which is bound and where most other lattice
results are centered and, the other close to where PDG
reports of $Z_b$ and $Z_b^\prime$ are. The experimentally observed states
are believed to contain a $b\bar{b}$ pair but ours is $bb$ pair which
is considered as theoretically simpler. However, the possible molecular
nature of $Z_b$ and $Z_b^\prime$ suggests that our molecular states can
have similar masses. For the bottom quarks we have used NRQCD action
while HISQ action for the $u/d$ quarks. This NRQCD-HISQ combination has
been employed earlier in \cite{eric} for bottom meson and recently in
\cite{protick} for bottom baryons. We have constructed the three
lattice $1^+$ trial states: a bound $|D\rangle$ containing ``good diquark" 
$3_c$ configuration and two meson-meson molecular $|M_1\rangle,\,
|M_2\rangle$ with the expectation that they will contribute to the
states above $B-B^*$ threshold. There are not many lattice results
on the states above
threshold possibly because of the complication that they can couple
to multiple decay channels besides $B^*\bar{B}$ and $B^*\bar{B}^*$.
Our motivation here is to obtain a tentative estimate of the $|M_1\rangle,
\,|M_2\rangle$ states above threshold and their relative overlap with the
bound state below the threshold.

An important component of the present investigation is the tuning of
the light $u/d$ quark mass. Depending on the wave function
of the operators we need two different tuning of the $u/d$ mass.
For the operators made of heavy-light $[b\bar{l}]$
mesonic wave functions we find it is necessary
to tune $am_l$ to match $b\bar{l}$ meson observed mass. Similarly,
for light-light diquark $[{l}_1{l}_2]$, where $l_1$ and $l_2$ may
or may not be equal, in presence of one or more heavy $b$ quarks
the $am_{l_i}$ is tuned with $\Lambda_b$. We applied this approach
with fair success in bottom baryon \cite{protick} and presently with
double bottom tetraquark we attempted the same. In order to understand
and explain these two different tuning, we solve the quantum mechanical
Hamiltonians of $B$-meson system, where a single light antiquark is in the
potential of a static bottom quark, and the $\Lambda_b$-baryon system,
where the two light quarks are in the same field of the static $b$
quark. In this problem $b$ mass is the experimental mass and the
light quark mass is treated as a parameter which is tuned to
reproduce the experimental $B$ and $\Lambda_b$ masses. We find that
the meson and baryon systems are solved for two different light quark
masses which justifies our need for two different tunings. However,
the actual numbers from these two sets of light quark masses, one
from solving the Schr\"{o}dinger equation and the other lattice tuned,
cannot be compared directly due to two different $b$ masses used in
these two instances.

Once tuned, we find the spectra of the lattice states $|D\rangle,\,
|M_1\rangle$ and $|M_2\rangle$. Naive calculation of $\langle \mathcal{O}_D
\mathcal{O}_D^\dagger \rangle$ spectrum yields a bound state
$-167\,(19)$ MeV measured from the $B-B^*$ threshold. On lattice,
states having the same quantum numbers can mix and,
therefore, it is natural to construct correlation matrices to
solve the generalized eigenvalue problem in order to obtain the
first few lowest lying energies. Besides, the components
of the eigenvectors provide the relative contribution of the trial
states, corresponding to the lattice operators, to the energy
eigenstates. They are the coefficient of expansion of the eigenstates
when expressed in terms of trial states as shown in the equation
(\ref{energybasis}). The GEVP analysis reveals that tetraquark
molecular state just above the threshold by only 17(14) MeV is
dominated by $|M_1\rangle$ lattice state while the lowest lying bound
state receives dominant contribution from $|D\rangle$ along with
significantly large contribution from both $|M_1\rangle$ and
$|M_2\rangle$.
From $3 \times 3$ $\Lambda_b$ tuned correlation matrix, we get
our final binding energy number for $bb\bar{u}\bar{d}$ tetraquark
system to be $-189\,(18)$ MeV, where the error is statistical.

\section{Acknowledgement}
The numerical part of this work, involving generation of heavy quark
propagators, has been performed at HPC facility in ``Kalinga" cluster
at NISER funded by Dept. of Atomic Energy (DAE), Govt. of India. The
construction of the correlators and other analysis part of this paper
has been carried out in the ``Proton" cluster funded by DST-SERB project
number SR/S2/HEP-0025/2010. The authors acknowledge useful discussions
with Rabeet Singh (Banaras Hindu University, India) on Hartree-Fock
calculation. One of the authors (PM) thanks DAE for financial support.



\section*{References}
\begin{thebibliography}{99}
\bibitem{bondar1} A.E. Bondar, A. Garmash, A.I. Milstein, R. Mizuk and
M.B. Voloshin, Phys. Rev. D {\bf 84}, 054010 (2011).
\bibitem{bondar2} A. Bondar {\it et al.} (Belle Collaboration), Phys.
Rev. Lett. {\bf 108}, 122001 (2012).
\bibitem{aaij} R. Aaij {\it et al.} (LHCb Collaboration), Phys. Rev.
Lett {\bf 112}, 222002 (2014).
\bibitem{lebed} R.F. Lebed, R.E. Mitchell and E.S. Swanson, Prog.
Part. Nucl. Phys. {\bf 93}, 143 (2017).
\bibitem{esposito} A. Esposito, A. Pilloni and A.D. Polosa, Phys. Rep.
{\bf 668}, 1 (2017).
\bibitem{olsen} S.L. Olsen, T. Skwarnicki and D. Zieminska, Rev. Mod.
Phys. {\bf 90}, 015003 (2018).
\bibitem{manohar} A.V. Manohar and M.B. Wise, Nucl. Phys. {\bf B399},
17 (1993).
\bibitem{eichten} E.J. Eichten and C. Quigg, Phys. Rev. Lett. {\bf 119},
202002 (2017).
\bibitem{ikeda} Y. Ikeda, B. Charron, S. Aoki, T. Doi, T. Hatsuda, T.
Inoue, N. Ishii, K. Murano, H. Nemura and K. Sasaki, Phys. Lett. {\bf
B729}, 85 (2014).
\bibitem{wagner} Marc Wagner {\it et al.} 2014 J. Phys.: Conf. Ser.
{\bf 503}, 012031.
\bibitem{padmanath} M. Padmanath, C.B. Lang and S. Prelovsek, Phys. Rev.
D {\bf 92}, 034501 (2015).
\bibitem{alexandrou} C. Alexandrou, J. Berlin, J. Finkenrath, T.
Leontiou and M. Wagner, Phys. Rev. D {\bf 101}, 034502 (2020).
\bibitem{sasa} S. Prelovsek, H. Bahtiyar and J. Petkovic, Phys. Lett.
B 805 (2020) 135467.
\bibitem{hughes} C. Hughes, E. Eichten and C.T.H. Davies, Phys. Rev.
D {\bf 97}, 054505 (2018).
\bibitem{francis1} A. Francis, R.J. Hudspith, R. Lewis and K. Maltman,
Phys. Rev. Lett. {\bf 118}, 142001 (2017)
\bibitem{francis2} A. Francis, R.J. Hudspith, R. Lewis and K. Maltman,
Phys. Rev. D {\bf 99}, 054505 (2019).
\bibitem{junnarkar} P. Junnarkar, N. Mathur and M. Padmanath, Phys.
Rev. D {\bf 99}, 034507 (2019).
\bibitem{luka} L. Leskovec, S. Meinel, M. Pflaumer and M. Wagner,
Phys. Rev. D {\bf 100}, 014503 (2019).
\bibitem{richard} J. P. Ader, J. M. Richard, P. Taxil, Phys. Rev. D
{\bf 25}, 2370 (1982).
\bibitem{bicudo1} P. Bicudo, and M. Wagner, Phys. Rev. D {\bf 87}, 
114511 (2013).
\bibitem{bicudo2} P. Bicudo, K. Cichy, A. Peters, B. Wagenbach and M.
Wagner, Phys. Rev. D {\bf 92}, 014507 (2015).
\bibitem{bicudo3} P. Bicudo, J. Scheunert and M. Wagner, Phys. Rev.
D {\bf 95}, 034502, (2017).
\bibitem{bicudo4} P. Bicudo, M. Cardoso, A. Peters, M. Pflaumer and
M. Wagner, Phys. Rev. D {\bf 96}, 054510 (2017).
\bibitem{thacker} B.A. Thacker and G.P. Lepage, Phys. Rev. D {\bf 43},
196 (1991).
\bibitem{lepage} G.P. Lepage, L. Magnea, C. Nakhleh, U. Magnea and K.
Hornbostel, Phys. Rev. D {\bf 46}, 4052 (1992).
\bibitem{hisq} E. Follana, Q. Mason, C. Davies, K. Hornbostel, G.P.
Lepage, J. Shigemitsu, H. Trottier and K. Wong, Phys. Rev. D {\bf 75},
054502 (2007).
\bibitem{fermilab} A.X. El-Khadra, A.S. Kronfeld and P.B. Mackenzie,
Phys. Rev. D {\bf 55}, 3933 (1997).
\bibitem{eric} E.B. Gregory {\it et al.} (HPQCD Collaboration), Phys.
Rev. D {\bf 83}, 014506 (2011).
\bibitem{kawamoto} N. Kawamoto and J. Smit, Nucl. Phys. B {\bf 192},
100 (1981).
\bibitem{ttqop} J. Jiang, W. Chen and S. Zhu, Phys. Rev. D {\bf 96}, 
094022 (2017).
\bibitem{jaffe} R.L. Jaffe, Phys. Rep. {\bf 409}, 1 (2005).
\bibitem{protick} P. Mohanta and S. Basak, Phys. Rev. D {\bf 101}, 
094503 (2020).
\bibitem{bazavov} A. Bazavov {\it et al.}, Rev. Mod. Phys. {\bf 82},
1349 (2010).
\bibitem{asqtad1} K. Orginos and D. Toussaint (MILC),
Phys. Rev. D {\bf 59}, 014501 (1998).
\bibitem{asqtad2} K. Orginos, D. Toussaint, and R. L. Sugar (MILC),
Phys. Rev. D {\bf 60}, 054503 (1999).
\bibitem{bowler_54} K. C. Bowler {\it et al.}, (UKQCD Collaboration),
 Phys. Rev. D {\bf 54}, 3619 (1996).
\bibitem{tiburzi} B.C. Tiburzi, Phys. Rev. D {\bf 71},
034501 (2005).
\bibitem{zachary} Z.S. Brown, W. Detmold, S. Meinel, and
K. Orginos, Phys. Rev. D90, 094507 (2014).
\bibitem{mes_q_mod} S. Godfrey and N. Isgur, Phys. Rev. D {\bf 32},
189 (1985).
\bibitem{bar_q_mod} S. Capstick and N. Isgur, Phys. Rev. D {\bf 34},
2809 (1986).
\bibitem{herman} \url{folk.uib.no/nfylk/Hartree/lindex.html}
\bibitem{hartree1} D. R. Hartree  and W. Hartree, Proc. R. Soc. Lond. 
A150: 9–33,  (1935).
\bibitem{hartree2} V. Fock,  Z. Physik 61, 126–148 (1930).
\bibitem{karliner} Marek Karliner and Jonathan L. Rosner, Phys. Rev.
Lett. {\bf 119}, 202001 (2017)
\bibitem{pdg} M. Tanabashi {\it et al.} (Particle Data Group), Phys. Rev.
D98, 030001 (2018) and 2019 update.
\bibitem{beane} S.R. Beane, K. Orginos, and M.J. Savage, Phys. Lett. B
{\bf 654}, 20 (2007).
\bibitem{iritani} T. Iritani {\it et al.} (HAL QCD Collaboration), JHEP
1610 (2016) 101.
\bibitem{yang} G. Yang, J. Ping, and J. Segovia, Phys. Rev. D {\bf 101},
014001 (2020)

\end{thebibliography}
\end{document}